\documentclass[epj]{svjour}
\usepackage{color}
\usepackage[normalem]{ulem}  % \sout{old text} for strikeout
\usepackage[dvipsnames]{xcolor}

\usepackage{graphics}
\begin{document}
\title{New equation of state involving Bose-Einstein condensate of antikaon 
for supernova and neutron star merger simulations}
\author{Tuhin Malik \inst{1}, Sarmistha Banik \inst{1}, \and 
Debades Bandyopadhyay\inst{2}
% etc
}                     % Do not remove
\mail{sarmistha.banik@hyderabad.bits-pilani.ac.in}
\institute{Birla Institute of Technology and Science, Pilani, Hyderabad Campus,
Hyderabad - 500078, India \and Saha Institute of Nuclear Physics, HBNI, 1/AF 
Bidhannagar, Kolkata-700064, India}
\date{Received: date / Revised version: date}
\abstract{We compute a new equation of state table including Bose-Einstein 
condensate of $K^{-}$ mesons for core collapse supernova and neutron star 
merger simulations. Nuclei and interacting nucleons in non-uniform matter is 
described in an extended version of the nuclear statistical equilibrium model 
including excluded volume effects whereas the uniform matter at higher 
densities is treated in the relativistic hadron field theory with density 
dependent couplings. The equation of state table is generated for a wide range
of density ($10^{-12}$ to $\sim 1$ fm$^{-3}$), positive charge fraction (0.01 to 0.60)
and temperature (0.1 to 158.48 MeV). The impact of antikaon condensate is
investigated on different thermodynamic quantities for example free energy per
baryon, entropy per baryon, pressure as well as compositions of matter. 
Furthermore, critical temperatures of antikaon condensation and
the phase diagram of matter are also studied in this article.         
\PACS{97.60.Jd neutron stars - 26.60.Kp equations of state
     } % end of PACS codes
} %end of abstract
\titlerunning{New Equation of State ...}
\authorrunning{T. Malik et al.}
\maketitle
\section{Introduction}
\label{intro}
Equation of state (EoS) of matter is an important microphysical input for 
core collapse supernova (CCSN) and binary neutron star (BNS) merger 
simulations. Many EoSs were developed based on updated knowledge derived from 
nuclear physics experimental data and neutron star observations  
\cite{hs1,horo,stei,toga,const,Sophia,Mamun,Blaschke,Sophia19,Sedra18,Sedra19,Raduta,Blaschke18}. 
One such widely used nuclear EoS known as Hempel and Schaffner (HS) EoS 
describes the inhomogeneous matter within the framework of the nuclear 
statistical equilibrium (NSE) model whereas the uniform dense nuclear matter is 
computed in the finite temperature relativistic mean field (RMF) model with 
and without density dependent (DD) couplings \cite{hs1,rmp}. 
The nuclear EoS table adopting the virial expansion 
for a non-ideal gas of nucleons and nuclei was computed by Shen et al. 
\cite{horo}. The nuclear EoS was also constructed in a variational
calculation using Argonne v18 (AV18) and Urbana IX (UIX) bare nuclear forces
\cite{toga,const}. The nuclear EoSs mentioned above are compatible with 
2 M$_{\odot}$ neutron stars. The first supernova EoS table that exploited 
measured masses and radii of neutron stars, was developed by Steiner and 
collaborators \cite{stei}.

Neutron star observations could provide important inputs in the construction of
EoS tables for CCSN and BNS merger simulations. Neutron star masses in 
relativistic binaries have been estimated to very high degree of precision 
owing to measurements of post-Keplerian parameters of pulsars such as orbital 
decay, periastron advance, Shapiro delay, time dilation.
Several heavy neutron stars of masses $\sim 2 M_{\odot}$ were 
discovered in the past decade. The first among those was 
PSR 0348+0432 having mass 2.01$\pm 0.04$ M$_{\odot}$ \cite{anto}. Now the
millisecond pulsar PSR 0740+6620 is credited as the most massive neutron star 
of 2.14$^{+0.10}_{-0.09}$ M$_{\odot}$ \cite{croma}. Recently Neutron Star 
Interior Composition Explorer (NICER) has reported the first simultaneous 
determination of mass and radius of PSR J0030+0451 \cite{watts}. This puts a 
stringent constraint on the $\beta$-equilibrated EoS of neutron star matter. 

It has been long debated that additional degrees of freedom in the form of
strange baryons\cite{weis}, quarks \cite{Blaschke18} or Bose-Einstein 
condensate of antikaons \cite{kaplan,knor} might appear in dense matter.
Many model calculations involving hyperon matter, quark
matter, Bose-Einstein condensate of antikaons, showed
that the EoS of $\beta$-equilibrated matter might result in 2 M$_{\odot}$ or 
more massive neutron stars \cite{weis,char14,dc}.
Several temperature dependent EoS tables including exotic matter 
such as quark and hyperon matter were constructed and employed in CCSN and BNS 
merger simulations 
\cite{ishi,naka08,irina,sumi,shen11,naka,oertel12,peres,bhb,char15,rad}. These 
EoS 
tables are functions of wide range of values of three parameters - baryon 
density, positive charge fraction and temperature. A 
few of those temperature dependent EoS tables with exotic matter such as 
Banik, Hempel and Bandyopadhyay $\Lambda$-hyperons EoS (BHB$\Lambda \phi$) 
where hyperon-hyperon interaction is mediated by $\phi$ mesons, was directly 
compatible with the 2 M$_{\odot}$ neutron star \cite {bhb}. 

So far, there is no EoS table involving Bose-Einstein condensate of antikaons
for CCSN and BNS merger simulations. This motivates us to construct an EoS
table with antikaon condensate. The paper is organised as follows. We
describe the EoS with an antikaon condensate within the NSE and field 
theoretical models in Section 2. Results are discussed in Section 3. We 
conclude in Section 4. 
\section{Equation of State}
The inhomogeneous matter is described by HS EoS \cite{hs1}. In this case, 
the ensemble of nuclei and nucleons are treated in an extended version of the 
NSE model using the RMF model for interacting nucleons. It also considers 
excluded volume effects in the thermodynamically consistent manner and excited 
states of light nuclei. The medium modification of nuclei due to screening of
Coulomb energies of background electrons are taken into account \cite{hs1}. 
Finally the sub-saturation density matter was matched with the dense uniform 
matter.

We adopt a density dependent relativistic hadron field theory for the 
description of strongly interacting dense baryonic matter. Here nucleon-nucleon
interaction is mediated by exchanges of scalar $\sigma$, vector $\omega$ and 
$\rho$ mesons. This is given by the Lagrangian density \cite{bhb,typ10}, 

\begin{eqnarray}
\label{eq_lag_b}
{\cal L}_{N} &=& \sum_{N} \bar\psi_{N}\left(i\gamma_\mu{\partial^\mu} - m_N
+ g_{\sigma N} \sigma - g_{\omega N} \gamma_\mu \omega^\mu \right. \nonumber\\
&& \left. - g_{\rho N} 
\gamma_\mu{\mbox{\boldmath $\tau$}}_N \cdot 
{\mbox{\boldmath $\rho$}}^\mu  \right)\psi_N\nonumber\\
&& + \frac{1}{2}\left( \partial_\mu \sigma\partial^\mu \sigma
- m_\sigma^2 \sigma^2\right)
-\frac{1}{4} \omega_{\mu\nu}\omega^{\mu\nu}\nonumber\\
&&+\frac{1}{2}m_\omega^2 \omega_\mu \omega^\mu
- \frac{1}{4}{\mbox {\boldmath $\rho$}}_{\mu\nu} \cdot
{\mbox {\boldmath $\rho$}}^{\mu\nu}
+ \frac{1}{2}m_\rho^2 {\mbox {\boldmath $\rho$}}_\mu \cdot
{\mbox {\boldmath $\rho$}}^\mu.
\label{had}
\end{eqnarray}
Here $\psi_N$ stands for nucleon doublet, ${\mbox{\boldmath 
$\tau_{N}$}}$ is the isospin operator and density dependent 
meson-nucleon couplings are denoted by $g_{\alpha N}$ with $\alpha$ representing
meson fields. 

\begin{figure*}
\begin{center}
%\vspace*{1cm}       % Give the correct figure height in cm
%\hspace{0.5cm}
\resizebox{1.0\textwidth}{!}{\includegraphics{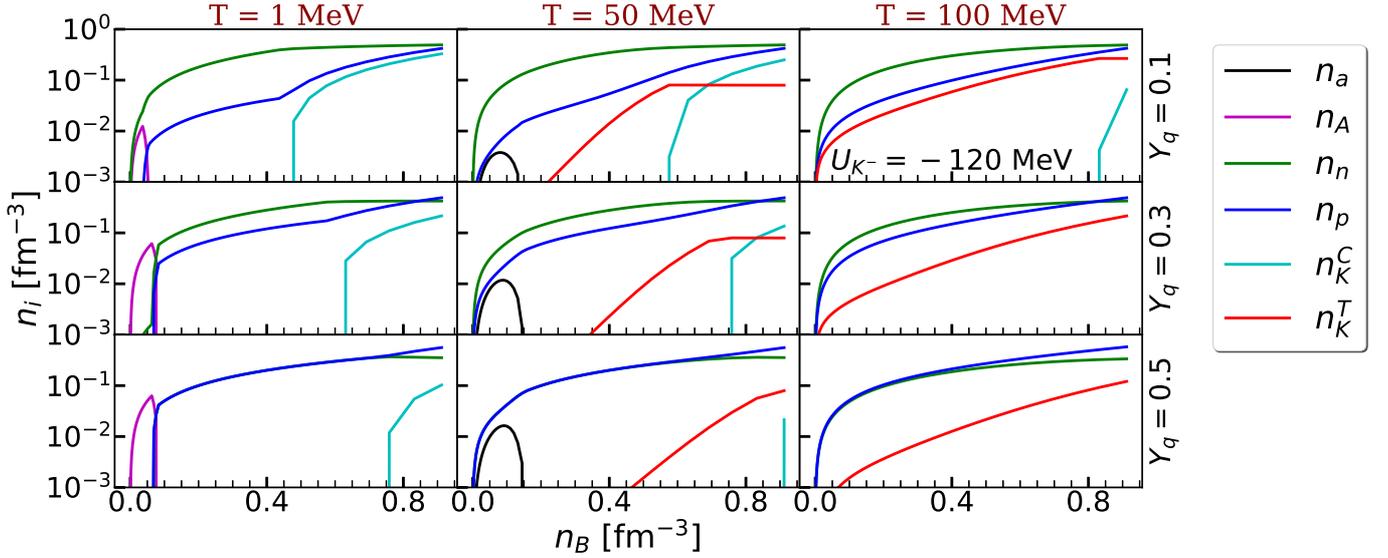}}
\caption{Number densities of different particle species, such 
as light \& heavy nuclei, neutrons, protons and antikaons, both thermal and 
condensate, as a function of baryon
number density for T=1, 50, 100 MeV and $Y_q$ = 0.1, 0.3, 0.5.}
\label{fig:1}       % Give a unique label
\end{center}
\end{figure*}
The grand-canonical thermodynamic potential per unit volume of the
nuclear phase is given by
\begin{eqnarray}
\frac{\Omega}{V} &=& \frac{1}{2}m_\sigma^2 \sigma^2
- \frac{1}{2} m_\omega^2 \omega_0^2 
- \frac{1}{2} m_\rho^2 \rho_{03}^2  
- \Sigma^r \sum_{i=n,p} n_i
\nonumber \\
&& - 2T \sum_{i=n,p} \int \frac{d^3 k}{(2\pi)^3} 
[\mathrm{ln}(1 + e^{-\beta(E^* - \nu_i)}) \nonumber\\ 
&& + \mathrm{ln}(1 + e^{-\beta(E^* + \nu_i)})] ~,  
\end{eqnarray}
where the temperature is defined as $\beta = 1/T$, effective nucleon mass
$m^{*}_{N} = m_N - g_{\sigma N} \sigma$ and $E^* = \sqrt{(k^2 + m_N^{*2})}$.
We calculate all thermodynamic quantities such as pressure $P = - {\Omega}/V$, 
energy density and entropy density. The chemical potential is given by
The nucleon chemical potential ($\mu_N$) is defined as
\begin{equation}
\mu_{N} = \nu_N + g_{\omega N} \omega_0 + g_{\rho N} \tau_{3N} \rho_{03}
+ \Sigma^r~,
\end{equation} 
where $\Sigma^r$ is the rearrangement term that takes care of many-body effects
in nuclear interaction \cite{bhb,typ10}.

%The onset of antikaon condensation occurs when the charge /electron chemical potential is equal to the in-medium energy of antikaons 
We consider a 
second order phase transition from the nuclear to antikaon condensed phase. 
Nucleons in the antikaon condensed phase behave differently than nucleons in 
the hadronic phase \cite{glen99}. 
Kaon-nucleon interaction is considered in the same footing as the 
nucleon-nucleon interaction in Eq. (\ref{had}). The Lagrangian density for 
(anti)kaons in the minimal coupling scheme is \cite{char14,glen99,banik01,sb08},
\begin{equation}
{\cal L}_K = D^*_\mu{\bar K} D^\mu K - m_K^{* 2} {\bar K} K ~,
\end{equation}
where $K$ and $\bar K$ denote kaon and (anti)kaon doublets; the covariant 
derivative is
$D_\mu = \partial_\mu + ig_{\omega K}{\omega_\mu} 
+ i g_{\rho K}
{\mbox{\boldmath t}}_K \cdot {\mbox{\boldmath $\rho$}}_\mu$ and
the effective mass of antikaons is $m_K^* = m_K - g_{\sigma K} \sigma$.

The thermodynamic potential for antikaons is given by \cite{sb08,pons},
\begin{equation}
\frac {\Omega_K}{V} = T \int \frac{d^3p}{(2\pi)^3} [ ln(1 - 
e^{-\beta(\omega_{K^-} - \mu)}) + 
 ln(1 - e^{-\beta(\omega_{K^+} + \mu)})]~.
\end{equation}
The in-medium energies of $K^{\pm}$ mesons are given by
\begin{equation}
\omega_{K^{\pm}} =  \sqrt {(p^2 + m_K^{*2})} \pm \left( g_{\omega K} \omega_0
+ \frac {1}{2} g_{\rho K} \rho_{03} \right)~,
\end{equation}
and $\mu$ is the chemical potential of $K^-$ mesons and is given by
$\mu = \mu_n -\mu_p$. The threshold condition for $K^-$ condensation is given
by
$\mu = \omega_{K^{-}} =   m_K^* - g_{\omega K} \omega_0
- \frac {1}{2} g_{\rho K} \rho_{03}$~.

The total (anti)kaon number density ($n_K$) made of thermal kaons ($n_K^{Th}$)
and the $K^-$ condensate ($n_K^C$), is given by 
$n_K = n_K^C + n_K^{T}~$, where 
\begin{eqnarray}
n^C_{K} &=& 2\left( \omega_{K^-} + g_{\omega K} \omega_0
+ \frac{1}{2} g_{\rho K} \rho_{03} \right) {\bar K} K
= 2m^*_K {\bar K} K  ~, \nonumber \\
n_K^{T}& = &
\int \frac{d^3 p}{(2\pi)^3}
\left({\frac{1}{e^{\beta(\omega_{K^-}-\mu)}
- 1}} - {\frac{1}{e^{\beta(\omega_{K^+}+\mu)} - 1}}\right)~. 
\end{eqnarray}
We calculate the pressure due to thermal kaons using $P_K = -{\Omega_K}/{V}$.
  
Meson field equations are solved in the mean field approximation. Meson fields are modified in the presence of $K^-$ condensate \cite{glen99,banik01}. Finally
we obtain the pressure versus energy density known as the EoS in nuclear and
antikaon condensed phases. For density dependent couplings, the pressure of 
nuclear matter includes the rearrangement term  for thermodynamic consistency. 
The functional forms of nucleon-meson couplings in baryon density are given by
Ref\cite{typ10}. The saturation density, mass of $\sigma$ meson, couplings at
the saturation density and unknowns of those functions are obtained by fitting
properties of finite nuclei. This parameter set is known as the DD2. Nuclear
matter properties at the saturation density are in consonance with those 
obtained from nuclear physics experiments \cite{rmp}. However, kaon-meson
couplings do not depend on density. In this case, kaon-vector meson coupling
constants are estimated exploiting the quark model and iso-spin counting rule
i.e. $g_{\omega K} = \frac{1}{3} g_{\omega N}$ and $g_{\rho K} = g_{\rho N}$
\cite{banik01,sch96}. The scalar coupling constant is determined from the real
part of $K^{-}$ optical potential at the saturation density 
($n_0=0.149065 fm^{-3}$), 
\begin{equation}
U_{K^{-}} = - g_{\sigma K}\sigma_0 - g_{\omega K} \omega_0 + {\Sigma}^{r}~.
\end{equation}
Although the study of kaonic atoms indicates the
an attractive $K^-$-nucleus potential, there is no consensus how deep it is.
The $K^-$ optical potential could range from $-60$ MeV to $-200$ MeV as follows
from
unitary chiral model calculations and phenomenological fit to kaonic atom data 
\cite{Fri94,Fri99,Tol,Tol2}.
For this calculation, we take an average antikaon optical 
potential depth U$_{K^{-}}$ = -120 MeV.
\section{Results and Discussion}
We generate the   EoS table with $K^-$ condensate using the 
DD2 parameter set and 
U$_{K^{-}}$ = - 120
MeV for a wide range of values of baryon density ($n_B$) ($10^{-12}$ to $\sim$ 
1 fm$^{-3}$), temperature (T) (0.1 to 158.48 MeV) and  positive
charge fraction ($Y_q$) defined as $Y_q n_B = n_p - n_K $,
(0.01 to 0.60). Grid spacings for baryon density is $\Delta$log$_{10}(n_B)$ = 
0.04; for temperature $\Delta$log$_{10}(T)$ = 0.04 and for  
positive charge fraction 
$\Delta Y_q$ = 0.01. There are 301 baryon density points, 81 temperatures and
 60 positive charge fractions in the EoS table. The matching 
of the non-uniform matter 
part of nucleonic HS(DD2) EoS \footnote{publicly available in
https://compose.obspm.fr/} \cite{fis14} with the kaon EoS is done in the 
following way. We merge two EoS tables and form a kaon EoS table 
including the non-uniform matter described by the extended NSE model. This 
merger is guided by the minimization of 
free energy per baryon at fixed $n_B$, T and $Y_q
$. 
We impose the condition that $n_{K}^T$ 
is $> 10^{-4}$ fm$^{-3}$ in the non-uniform matter in 
addition to the above physical criterion as the contribution of 
antikaons to thermodynamic observables is negligible for $n_{K}^T < 10^{-4}$
fm$^{-3}$. Henceforth 
we call this merged EoS table 
\footnote{https://universe.bits-pilani.ac.in/Hyderabad/sbanik/EoS} 
with $K^{-}$ as HS(DD2)K$^-$ .
\begin{figure*}
\begin{center}
\resizebox{1.0\textwidth}{!}{\includegraphics{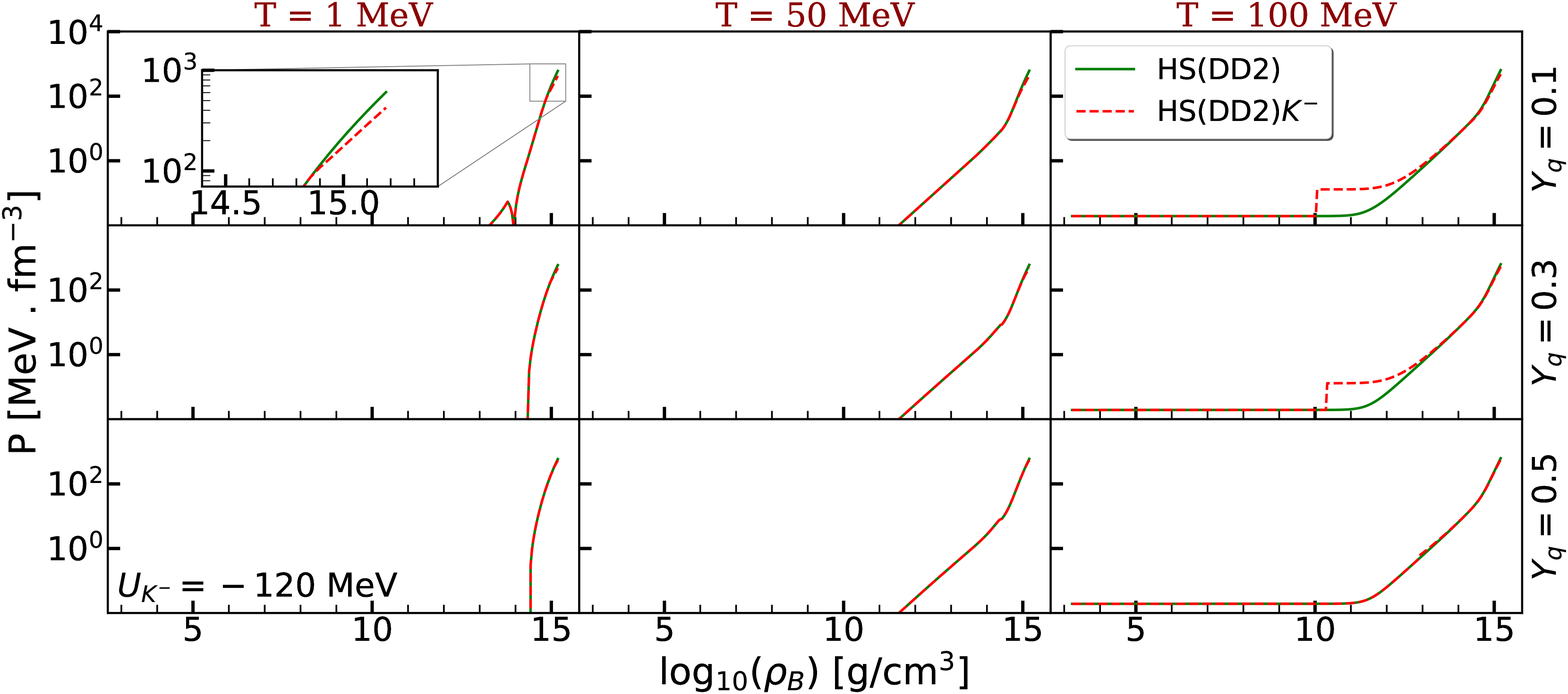}}
\caption{Pressure is shown as a function of baryon mass density for T=1, 50, 
100 MeV and $Y_q$ = 0.1, 0.3, 0.5. The difference in the EoS 
at higher density is exhibited in the inset for $Y_q$ = 0.1 and T=1 MeV.}
\label{fig:2}       % Give a unique label
\end{center}
\end{figure*}

Populations of particles are shown for T=1, 50, 100 MeV and $Y_q
$ = 0.1, 0.3, 
0.5 in Fig. 1. Light ($Z \leq 5$) and heavy ($Z \geq 6$) nuclei appear in 
non-uniform matter at temperatures T=1, 50 MeV. Number 
densities of light and heavy nuclei are denoted by $n_a$ and $n_A$, 
respectively. Nuclei dissolve into uniform matter of neutrons and protons
around the saturation density. It is evident from the left panel of T=1 MeV, 
no thermal (anti)kaons are populated in the 
system. The Bose-Einstein condensate of $K^-$ mesons sets in at higher baryon 
densities. It is noted that the density of $K^-$ mesons ($n_{K^-}^{C}$) in the 
condensate is much higher for lower values of positive charge fractions for example at 
$Y_q
$=0.1. Furthermore, the proton number density ($n_p$) increases with the 
onset of the 
antikaon condensate and even exceeds the neutron number density ($n_n$) for 
higher values of $Y_q
$. The middle panel of T=50 MeV exhibits similar features 
of T=1 MeV panel with the population of thermal $K^{-}$ mesons. However, the 
density of $K^-$ mesons in the condensate dominates over that of thermal $K^-$ 
mesons ($n_{K^-}^{Th}$) except for $Y_q
$=0.5. The right panel shows the 
significant populations of thermal $K^{-}$ mesons at T=100 MeV for all positive charge fractions. However, in this case, the antikaon condensate disappears except for 
$Y_q = 0.1$.         

We have studied various thermodynamic observables such as 
free energy per baryon, entropy per baryon and pressure. 
The pressure is plotted in Fig. 2, as a function of
baryon mass density. Results are shown for T=1, 50, 100 MeV and
$Y_q
=$0.1, 0.3, 0.5. In all these cases, hadronic contributions are considered.
The pressure does not include the contribution of leptons.
It is noted that results for HS(DD2) and HS(DD2)$K^-$ EoSs 
do not show any difference for T=1, 50 MeV and three positive charge
fractions, at the lower density regions. However, at the higher density 
the antikaon condensates appear, which clearly  
makes the HS(DD2)K$^-$ EoS softer 
compared to HS(DD2) EoS. The high density portion  is zoomed 
in the inset box for top-left panel. However we find that
there are jumps in pressure for T=100 MeV and $Y_q
=$0.1 and 0.3 due to 
significant contribution of thermal $K^-$ mesons to the pressure at 
$\sim 10^{10}$ g/cm$^3$. 
A smooth transition from nuclear matter to npK$^-$ matter is observed 
for temperatures below 80 MeV. 

\begin{figure}
\vspace*{-1cm}       % Give the correct figure height in cm
\resizebox{0.5\textwidth}{!}{\includegraphics{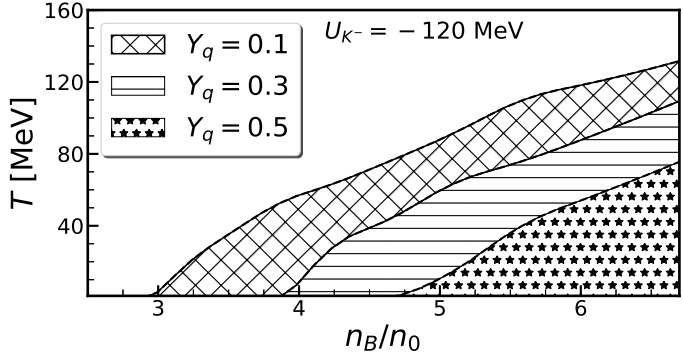}}
\caption{The phase diagram of nuclear matter with antikaon condensate is 
displayed for $Y_q
$ = 0.1, 0.3, 0.5 and U$_{K{^-}}= -120$ MeV.
The $K^-$ condensed phases, represented by the shaded regions are demarcated 
from the nuclear phases by the solid lines for the three $Y_q
$ values.
}
\label{fig:3}       % Give a unique label
\end{figure}
It is expected that the antikaon condensate would cease to exist above a 
critical temperature ($T_c$) \cite{sb08}. This critical temperature above which
the $K^-$ condensate dissolves, is a function of baryon density and positive 
charge fraction. We extract critical temperatures at different values of $n_B$ 
and $Y_q
$ and plot the phase diagram, $T$ versus $n_B$, in Fig.~\ref{fig:3} . Each line
in the plot denotes a particular value of $Y_q
$. 
Furthermore, the 
region below each line represents the $K^-$ condensate phase whereas the region
above it is the nuclear phase. It is observed that the antikaon condensate 
region shrinks as the value of $Y_q
$ increases.  
 
\begin{figure}
\begin{center}
\vspace{-0.4cm}       % Give the correct figure height in cm
%\hspace{0.5cm}
\resizebox{0.3\textheight}{!}{\includegraphics{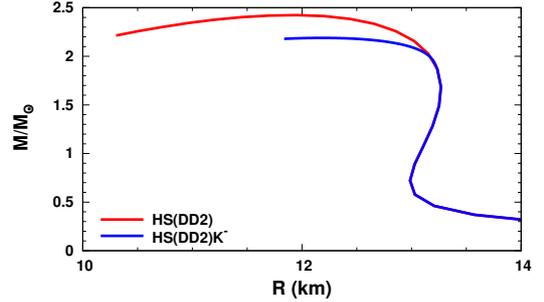}}
\caption{Mass versus radius of a charge neutral, $\beta$-equilibrated 
neutron star is displayed for HS(DD2) and HS(DD2)K$^-$ with
U$_{K^{-}}= -120$ MeV, at T=0.1MeV.}
\label{fig:4}       % Give a unique label
\end{center}
\end{figure}
Finally in Fig.~\ref{fig:4}, mass of neutron star is plotted with radius for charge 
neutral and 
$\beta$-equilibrated HS(DD2) and HS(DD2)K$^-$ EoSs at T=0.1 MeV. It is 
observed that the appearance of kaon condensate make the EoS softer resulting 
in lower maximum mass neutron star. Maximum mass and the corresponding radii
in two cases are 2.42 M$_{\odot}$, 11.95 km and 2.19 M$_{\odot}$, 12.14 km, 
respectively.
The tidal deformability for 1.4$M_{\odot}$ neutron 
star 
($70 \leq {\Lambda_{1.4}} \leq 580$) extracted from the BNS merger GW170817
provides radius of 1.4$M_{\odot}$ neutron star in the range $11.9^{+1.4}_{-1.4}$
km \cite{bpa}.
Similarly the NICER observation of PSRJ0030+0451 gives a radius of
$13.02^{+1.24}_{-1.06}$ km for the 1.44$M_{\odot}$ pulsar \cite{mil}. Our
results for radius corresponding to 1.4M$_{\odot}$ neutron star is slightly
higher than that of GW170817 \cite{soma} whereas it is consistent with the NICER
observation.
\section{Conclusions}
We have constructed a new EoS table including thermal $K^{-}$ mesons and the
Bose-Einstein condensate of $K^-$ mesons known as HS(DD2)$K^-$ for CCSN and BNS
merger simulations. In this work, an extended version of the NSE model with 
excluded volumes is adopted for the non-uniform matter whereas the uniform 
matter at higher densities is described in the relativistic hadron field theory
with density dependent couplings. The role of the antikaon condensate on 
various thermodynamic observables is investigated. The compositions of matter 
are appreciably modified due to the onset of the antikaon condensate at higher 
densities and thermal $K^{-}$ at higher temperatures. The presence of the 
antikaon condensate makes the HS(DD2)K$^-$ EoS softer than that of nucleonic 
HS(DD2) EoS. The phase diagram with the $K^-$ condensate is also 
investigated. Furthermore, the charge neutral and $\beta$-equilibrated
HS(DD2)K$^-$ EoS results in a lower maximum mass neutron star compared with 
that of HS(DD2) EoS. We plan to include hyperons along with the antikaon 
condensate in a future publication.  

\vspace{0.2cm}

\leftline {\bf Acknowledgements}
Authors acknowledge the DAE-BRNS grant received under the BRNS project 
No.37(3)/14/12/2018-BRNS.\\

\leftline{\bf Author Contribution}
All authors contribute equally to this work and agree to the published version 
of the manuscript.

\end{document}